\documentclass[twocolumn,twoside,slac_two]{revtex4}
\usepackage{graphicx}
\usepackage{fancyhdr}
\begin{document}

\title{Dark Matter implications of Fermi-LAT measurement of anisotropies in the
diffuse gamma-ray background}

%

\author{G.A. G\'omez-Vargas}
\affiliation{ Universidad Aut\'onoma de Madrid \& Instituto de F\'isica Te\'orica UAM/CSIC, Cantoblanco, E-28049 Madrid, Spain}
\affiliation{Istituto Nazionale di Fisica Nucleare, Sez. Roma Tor Vergata, Roma, Italy}
\author{A. Cuoco}
\affiliation{The Oskar Klein Centre for Cosmo Particle Physics, AlbaNova,
SE-106 91 Stockholm, Sweden}

\author{T. Linden}
\affiliation{Department of Physics, University of California, Santa Cruz, CA 95064, USA}

\author{M.A. S\'anchez-Conde}
\affiliation{SLAC National Accelerator Laboratory \& Kavli Institute for Particle Astrophysics and Cosmology, Menlo Park, CA 94025, USA}

\author{J.M. Siegal-Gaskins}
\affiliation{California Institute of Technology, Pasadena, CA 91125, USA\\}

\author{\textbf{For the Fermi-LAT Collaboration}\\}

\author{\\ T. Delahaye}
\affiliation{LAPTh, Universit e de Savoie, CNRS, 9 chemin de Bellevue, BP110, F-74941 Annecy-le-Vieux Cedex, France}
\affiliation{Institut d'Astrophysique de Paris, UMR 7095 - CNRS, Universit e Pierre \& Marie Curie, 98 bis boulevard Arago, 75014, Paris, France}
\affiliation{Instituto de F\'isica Te\'orica UAM/CSIC, Cantoblanco, E-28049 Madrid, Spain}

\author{M. Fornasa}
\affiliation{School of Physics and Astronomy, University of Nottingham, NG7 2RD Nottingham, United Kingdom}

\author{E. Komatsu}
\affiliation{Max-Planck-Institut fur Astrophysik, Karl-Schwarzschild Str. 1, 85741 Garching, Germany}
\affiliation{Kavli Institute for the Physics and Mathematics of the Universe, Todai Institutes for Advanced Study, the University of Tokyo, Kashiwa, Japan 277-8583 (Kavli IPMU, WPI)}
\affiliation{Texas Cosmology Center and the Department of Astronomy, The University of Texas at
Austin, 1 University Station, C1400, Austin, TX 78712, USA}

\author{F. Prada}
\affiliation{Campus of International Excellence UAM+CSIC, Cantoblanco, E-28049 Madrid, Spain}
\affiliation{Instituto de F\'isica Te\'orica UAM/CSIC, Cantoblanco, E-28049 Madrid, Spain}
\affiliation{Instituto de Astrof\'isica de Andaluc\'ia (CSIC), Glorieta de la Astronom\'ia, E-18080 Granada, Spain}

\author{J. Zavala}
\affiliation{Perimeter Institute for Theoretical Physics, 31 Caroline St. N., Waterloo, ON, N2L 2Y5, Canada}
\affiliation{Department of Physics and Astronomy, University of Waterloo, Waterloo, Ontario, N2L 3G1, Canada}

\begin{abstract}
The detailed origin of the diffuse gamma-ray background is still unknown. However, the contribution of
unresolved sources is expected to induce small-scale anisotropies in this emission, which may provide a way
to identify and constrain the properties of its contributors. Recent studies have predicted the contributions to
the angular power spectrum (APS) from extragalactic and galactic dark matter (DM) annihilation or decay. The Fermi-LAT collaboration reported detection of angular power with a significance larger than $3\sigma$ in the energy range from $1$ GeV to $10$ GeV on 22 months of data~\citep{2012PhRvD..85h3007A}. For these preliminary results the already published Fermi-LAT APS measurements~\citep{2012PhRvD..85h3007A} are compared to the accurate predictions for DM anisotropies from state-of-the-art cosmological simulations as presented in~\citep{2013MNRAS.429.1529F} to derive constraints on different DM candidates.
\end{abstract}

\maketitle
\thispagestyle{fancy}

\section{THE DIFFUSE GAMMA-RAY BACKGROUND}
\begin{figure*}[ht]
\centering
\includegraphics[width=100mm]{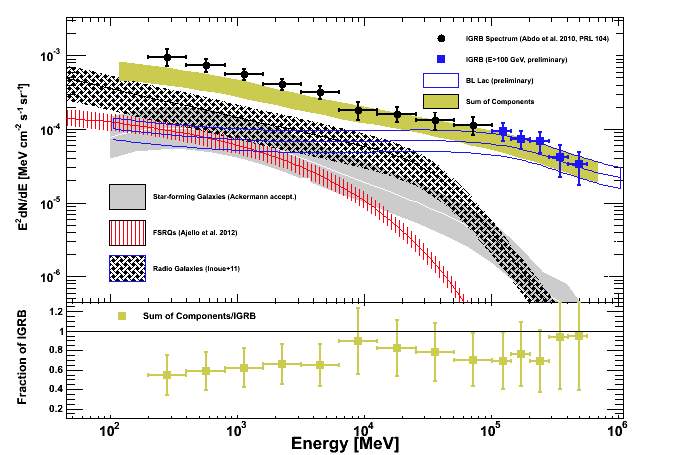}
\caption{IGRB spectrum and the contribution from all the different extragalactic source classes that have been detected by Fermi. The golden band shows the sum of all the source populations. By summing all contributions it is apparent that there is still room for other components at all energies within the uncertainties~\citep{2012AAS...22011605S}. See also~\citep{2011ApJ...736...40S} for other theoretical estimates of the relative contributions of unresolved blazars and star-forming galaxies to the IGRB.} \label{marco}
\end{figure*}

The diffuse gamma-ray background is characterized by an isotropic or nearly isotropic distribution and is therefore
known as the Isotropic Gamma-ray Background (IGRB)~\citep{1972ApJ...177..341K}.
It is constituted by gamma rays produced by various sources, including blazars, pulsars, and possible DM structures, not yet detected due to the limited angular resolution and photon statistics of the Fermi-LAT.
Figure \ref{marco} shows the IGRB spectrum and the estimated contributions from unresolved blazars, star-forming
and radio galaxies. The angular distribution of photons in the diffuse background may contain information about the
presence and the nature of these unresolved source populations.

\section{FERMI-LAT MEASUREMENT OF THE ANGULAR POWER SPECTRUM IN THE IGRB}

In ~\citep{2012PhRvD..85h3007A} the first 22 months of Fermi-LAT data were analyzed, dividing the energy
range between $1$ GeV and $50$ GeV in $4$ energy bins. The point sources in the first year catalogue \citep{2010ApJS..188..405A} have been masked, as well as the emission within a band of 30 degrees above and below the Galactic plane
. The masking was done
to cover the regions in the sky where the emission is dominated by resolved sources and by the Galactic foreground,
and to restrict the analysis only to where the IGRB is a significant component.

Two definitions for anisotropies were used in~\citep{2012PhRvD..85h3007A}:
\begin{itemize}
  \item Intensity APS: An intensity map $I(\psi)$ can be decomposed in spherical harmonics,
  \begin{equation}
    I(\psi)=\sum_{lm}{a_{lm}Y_{lm}(\psi)},
  \end{equation}
  where coefficients $a_{lm}$ determine the APS which is given by $C_{l}=\left< \left| a_{lm}\right|^2 \right>$.
  This definition is particularly useful because it gives us the dimensionful size of intensity fluctuations and can be compared with predictions for source classes whose collective intensity is known or assumed.
  \item Fluctuation APS: can be derived from the intensity APS, dividing by the average intensity squared. The fluctuation APS is energy-independent for a single source class, if all members of the source class share the same observed energy spectrum.
\end{itemize}

The Fermi-LAT collaboration reported detection of angular power in all 4 energy bins considered, with a signiﬁcance larger than $3\sigma$ in the energy bins from $1$ GeV to $10$ GeV. The data have been compared with the APS of a source model made of i) the point sources in~\citep{2010ApJS..188..405A}, ii) a model for the interstellar diffuse emission and iii) an isotropic component at the level of the IGRB in~\citep{2010PhRvL.104j1101A}. The model angular power at $155 \leq \textit{l} \leq 504$ is consistently below that measured in the data. 

Despite the mask applied along the Galactic plane, some known Galactic emission can extend to high latitudes. Therefore a model of the Galactic foregrounds was subtracted from the data, and then the APS of the residual maps was calculated. This measurement is referred to as the cleaned data in~\citep{2012PhRvD..85h3007A}. We use this second measurement in this work.

\section{DARK MATTER PREDICTIONS}
The APS of gamma rays from DM annihilations or
decays has been computed from the all-sky
template maps produced in~\citep{2013MNRAS.429.1529F}.
The authors of~\citep{2013MNRAS.429.1529F} used the Millennium-II N-body
simulation to model the abundance and the
clustering of extragalactic DM halos and subhalos.
The technique presented in~\citep{4}, based on the
random repetition of copies of the Millennium-II
simulation box, is implemented to probe the
universe up to $z=2$. The emission from DM halos
with a mass below the resolution of the
simulation was estimated assuming that the halo
number density and the mass-luminosity relation
obtained from the halos in Millennium-II remains
unchanged below the resolution, down to the
minimal self-bound halo mass $M_{min}$. On the other
hand, the contribution of low mass subhalos was
modeled following the technique described in~\citep{5}
and extended in~\citep{6}.

The smooth DM halo of the Milky Way was
parametrized in~\citep{2013MNRAS.429.1529F} as an Einasto profile, since
this provides the best fit to the Milky Way-like halo
obtained in the Aquarius N-body simulation.
Galactic subhalos down to $10^5$ $M_\odot$ were
accounted for directly from the Aquarius simulation,
while we use the same procedure as before to account for the contribution of
unresolved subhalos. It has been shown that such
objects do not contribute significantly to the total
intensity APS.

In \citep{2013MNRAS.429.1529F} the effect of the
assumptions made in the modeling of the DM
distribution also was estimated, looking for their effect both on the
intensity of the DM-induced emission and on its
APS. The two most relevant sources of
uncertainty are the amount of substructures
hosted by DM halos and the value of $M_{min}$. The first gives rise to an uncertainty of a factor 20-30
both in the average intensity and in the intensity
APS; when the uncertainty on the value of $M_{min}$ is
taken into account, these factors build up to $40$
and $100$ for the average intensity and the intensity
APS, respectively.

\section{SETTING CONSTRAINTS}\label{setting}
Here we present the method used to set conservative limits on the thermally averaged cross section, $\left<\sigma v\right>$, for DM annihilation into three different channels, $b\bar{b}$ quarks, $\mu^+\mu^-$, and $\tau^+\tau^-$ leptons.
\begin{itemize}
 \item To set constraints we use the foreground-cleaned $C_p$ shown in table II of~\citep{2012PhRvD..85h3007A}. There are four $C_p$ values measured corresponding to four energy bins, $1-2$ GeV, $2-5$ GeV, $5-10$ GeV, and $10-50$ GeV. We use them independently to set limits. 
 \item Requiring that the DM-induced intensity APS averaged in $155<l<204$ does not overshoot  the measured $C_p$ in the $155<l<504$ multipole range plus $1.64$ times its error leads to $95\%$ CL limits on  $\left<\sigma v\right>$\footnote{We assume Gaussian erros, then the value of $1.64$ is based on the fact that $95\%$ of the area of a Gaussian distribution is within $1.64$ standard deviations of the mean.}.
 \item We know that the IGRB anisotropy has multiple contributors, therefore these constraints are conservative. Other
contributors to IGRB anisotropy are not well known, but we already have constraints on the contribution of blazars~\citep{9}. We subtract this contribution from the measured $C_p$ and require that the DM-induced APS does not overshoot this new
limit.
\end{itemize}

\section{PRELIMINARY RESULTS}
\begin{figure*}[!t]
\centering
\includegraphics[width=75mm]{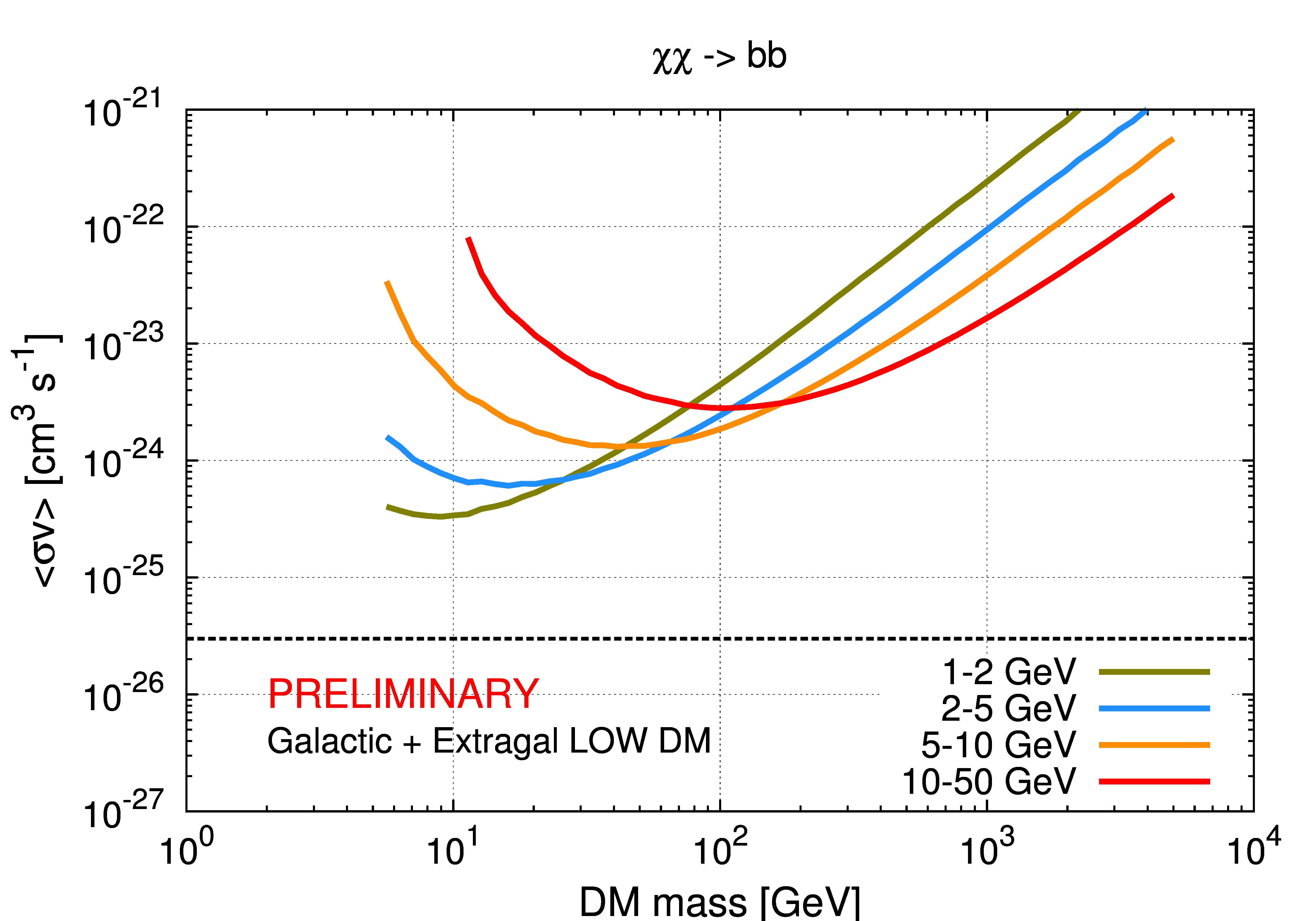}
\includegraphics[width=75mm]{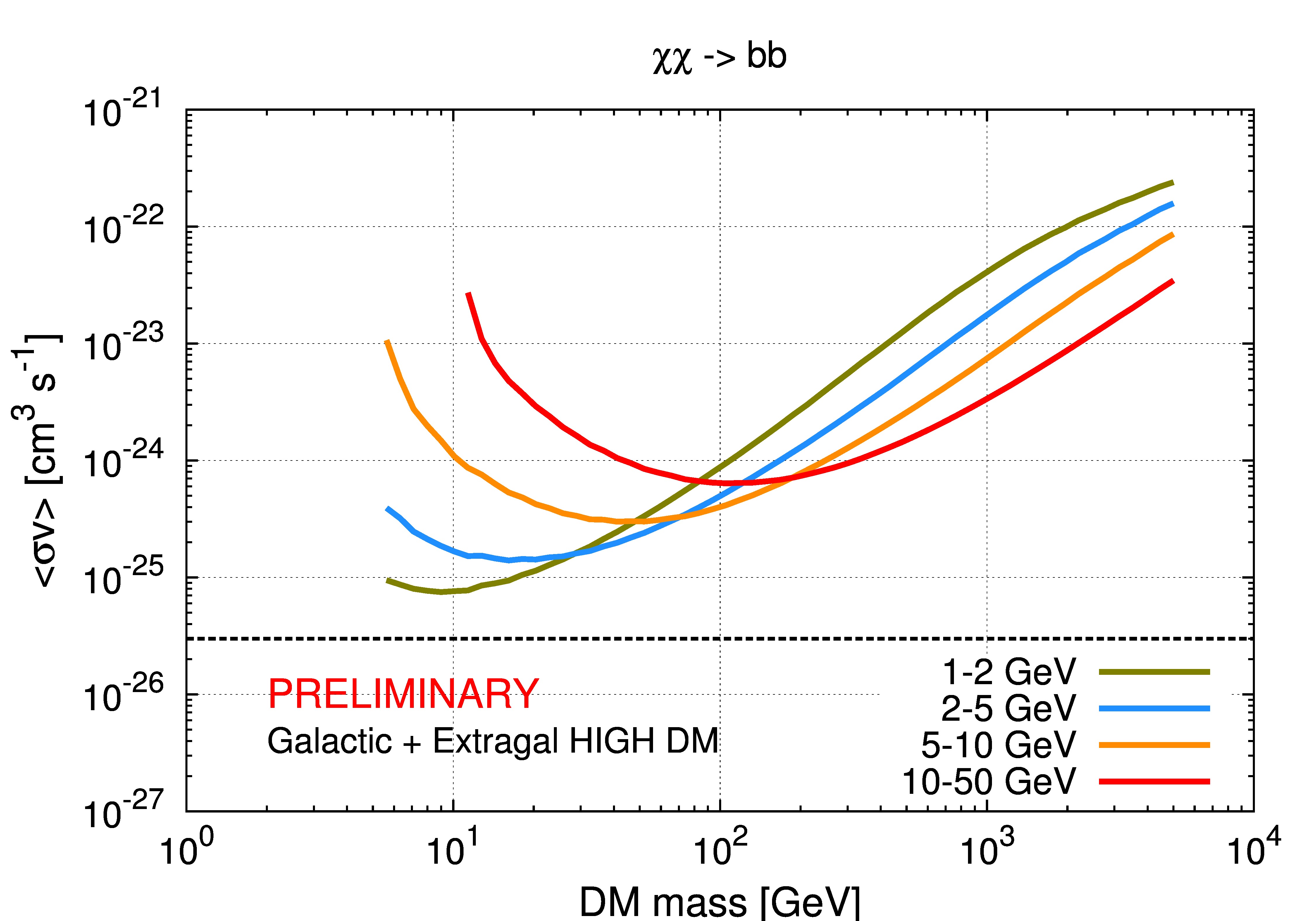}
\includegraphics[width=75mm]{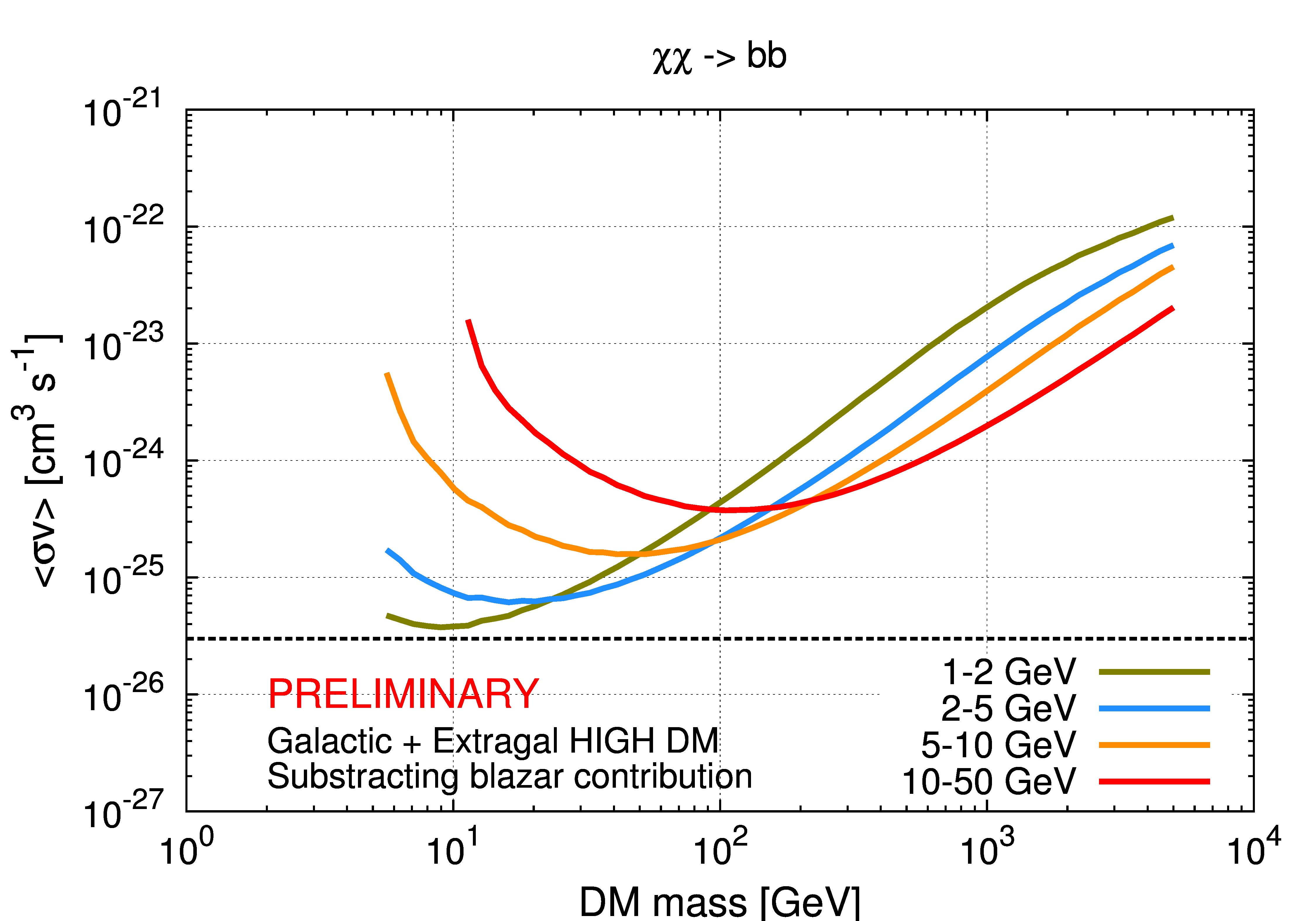}
\caption{$95\%$ CL limits on the annihilation cross section for the $b\bar{b}$ channel.} \label{bb}
\end{figure*}

\begin{figure*}[!t]
\centering
\includegraphics[width=75mm]{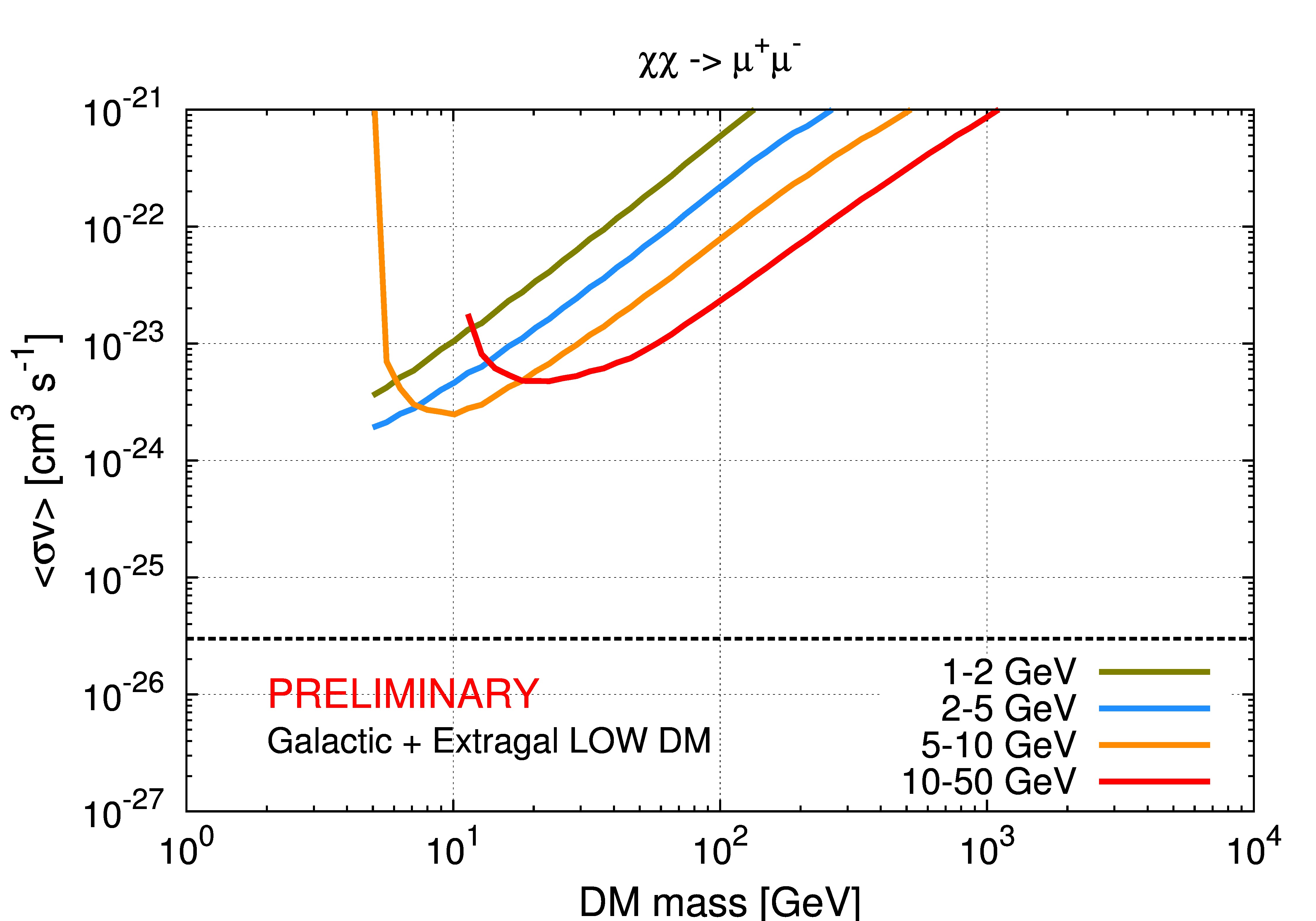}
\includegraphics[width=75mm]{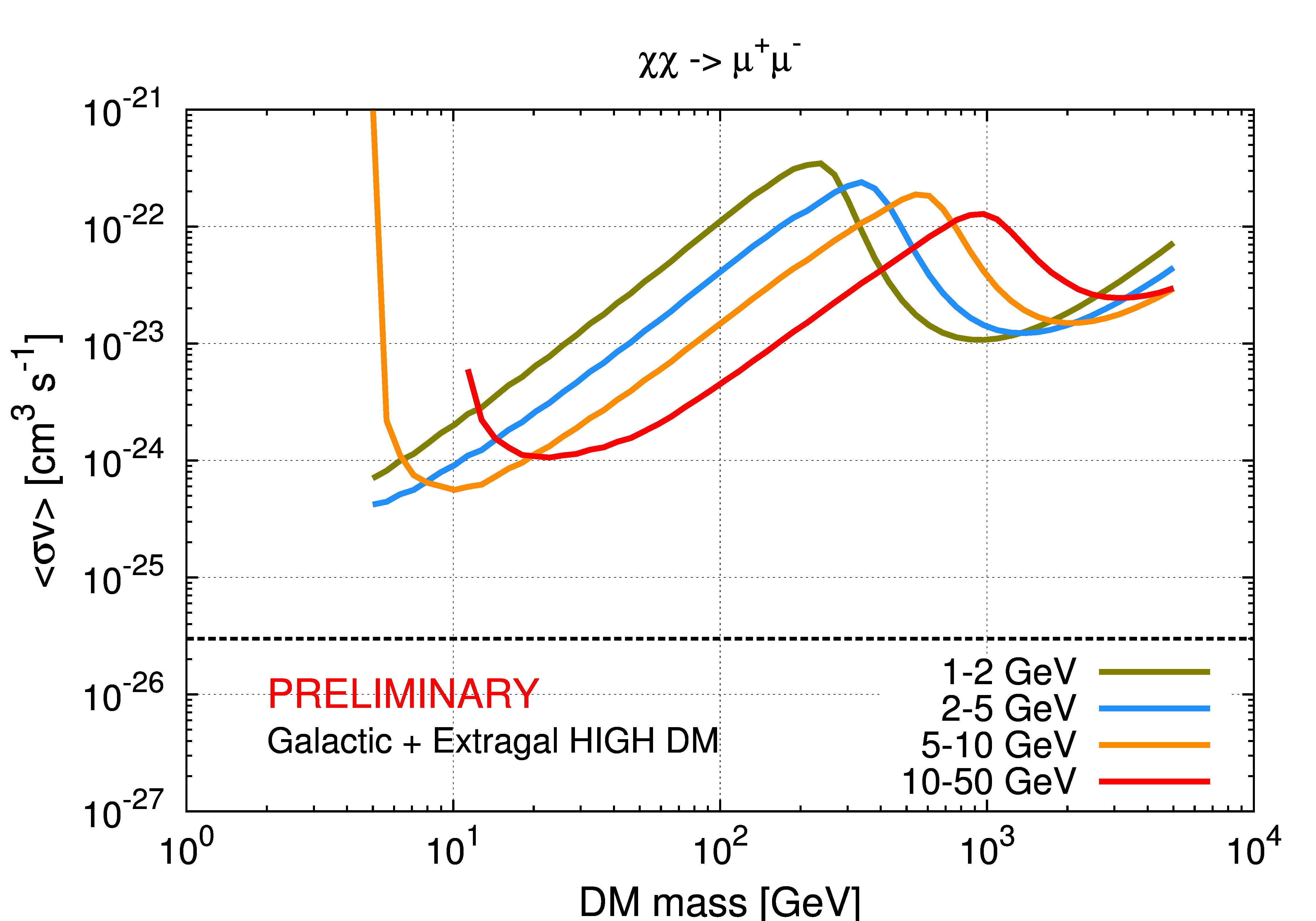}
\includegraphics[width=75mm]{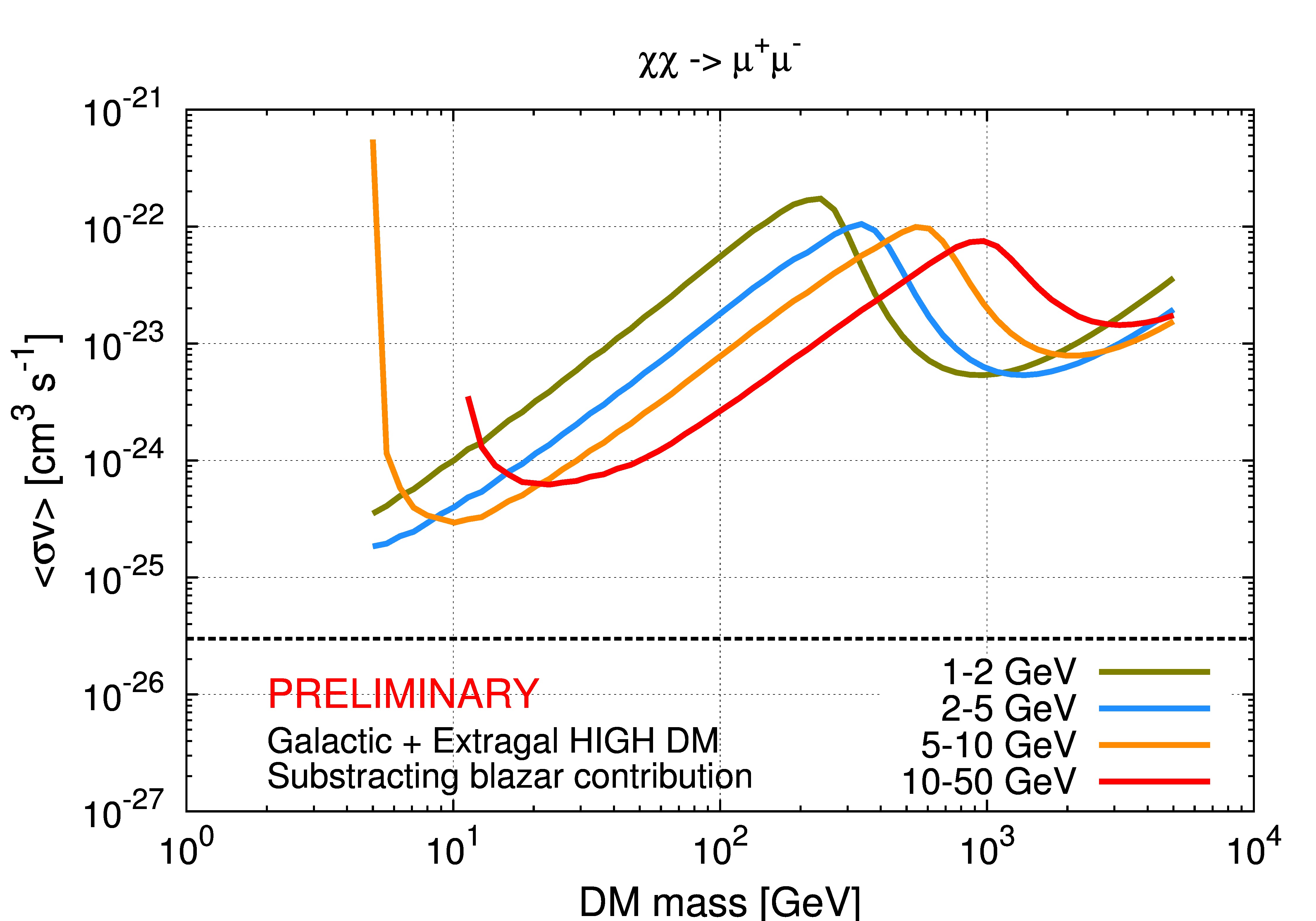}
\caption{$95\%$ CL limits on the annihilation cross section for the $\mu^+\mu^-$ channel.} \label{mu}
\end{figure*}

\begin{figure*}[!t]
\centering
\includegraphics[width=75mm]{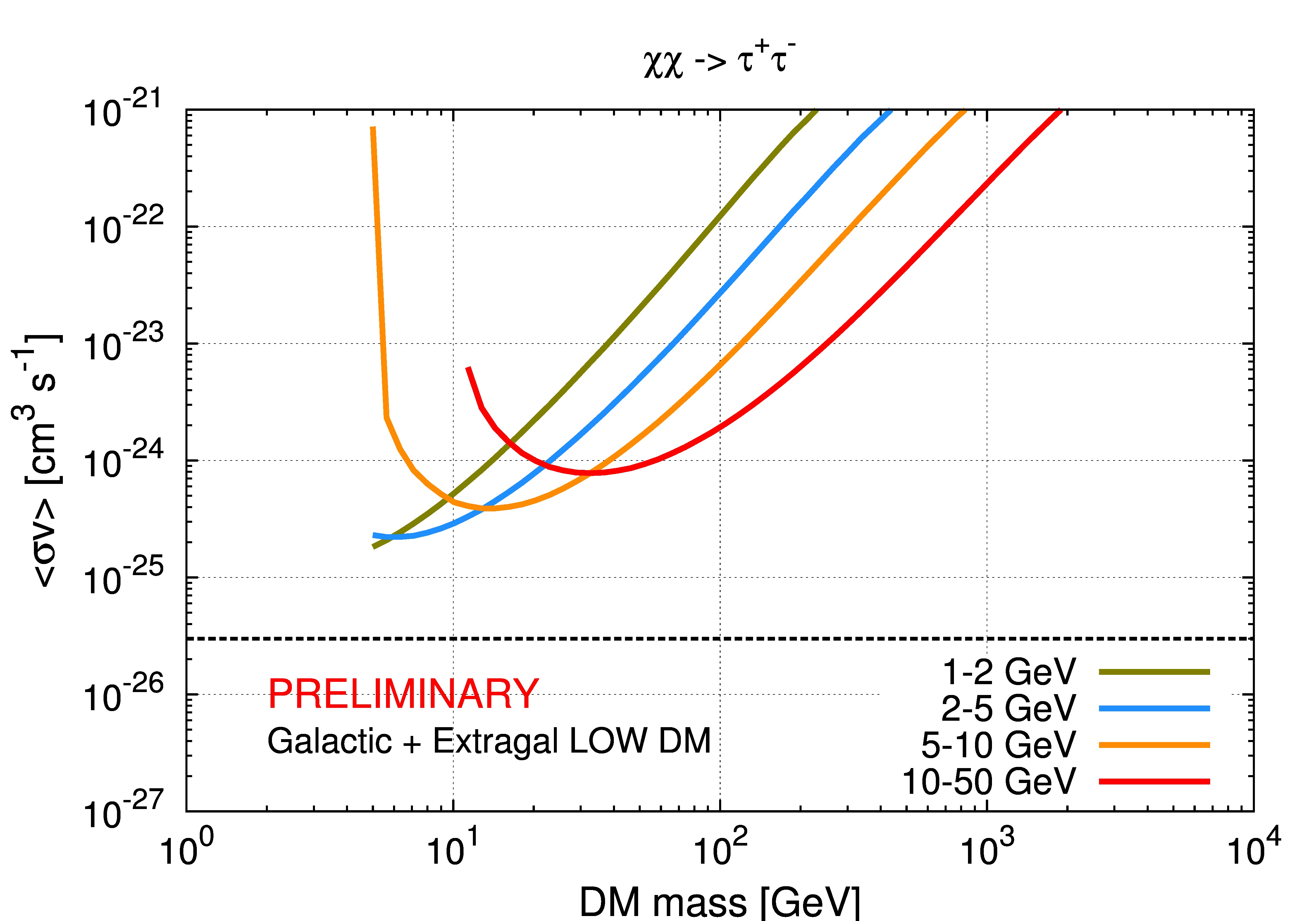}
\includegraphics[width=75mm]{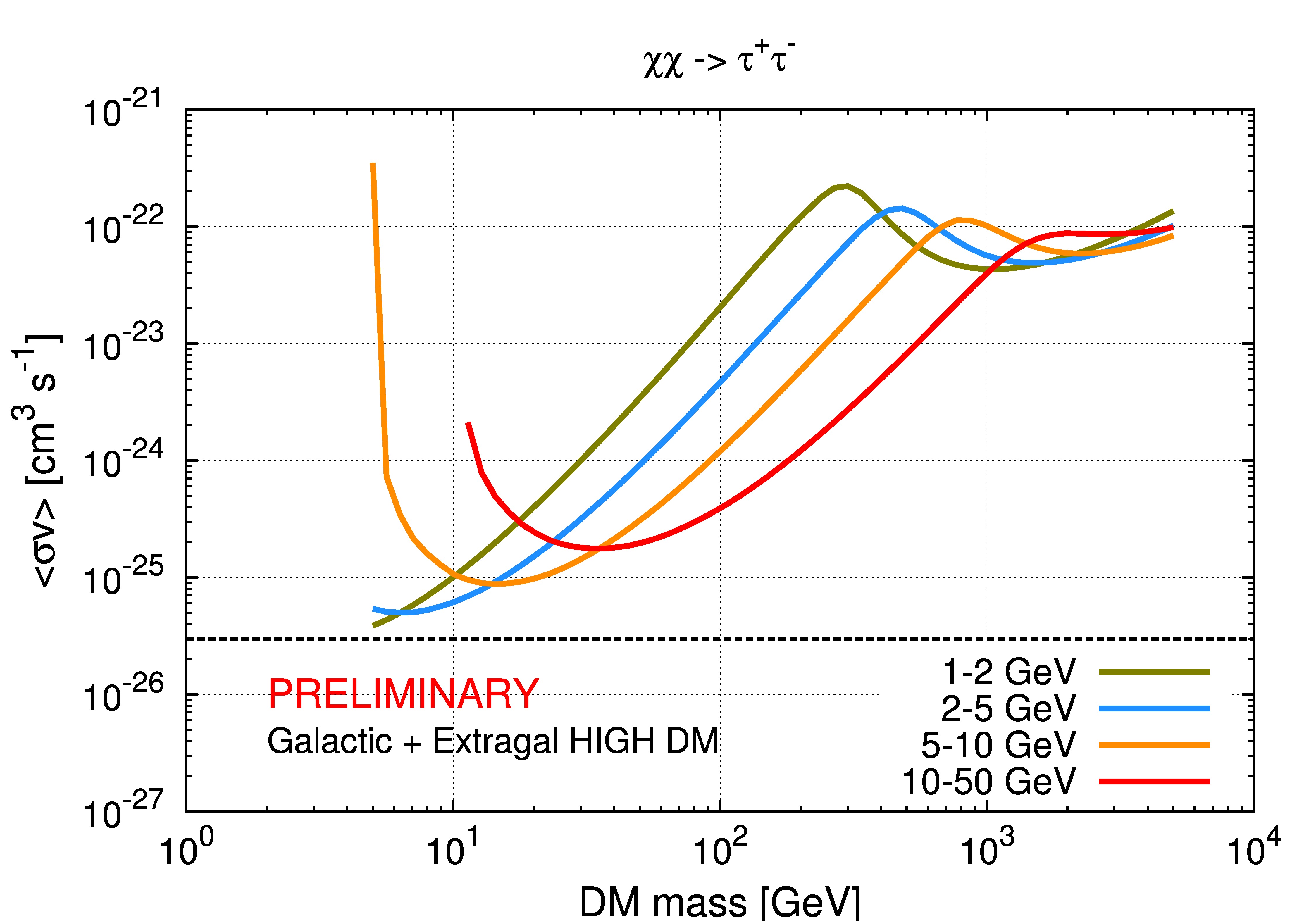}
\includegraphics[width=75mm]{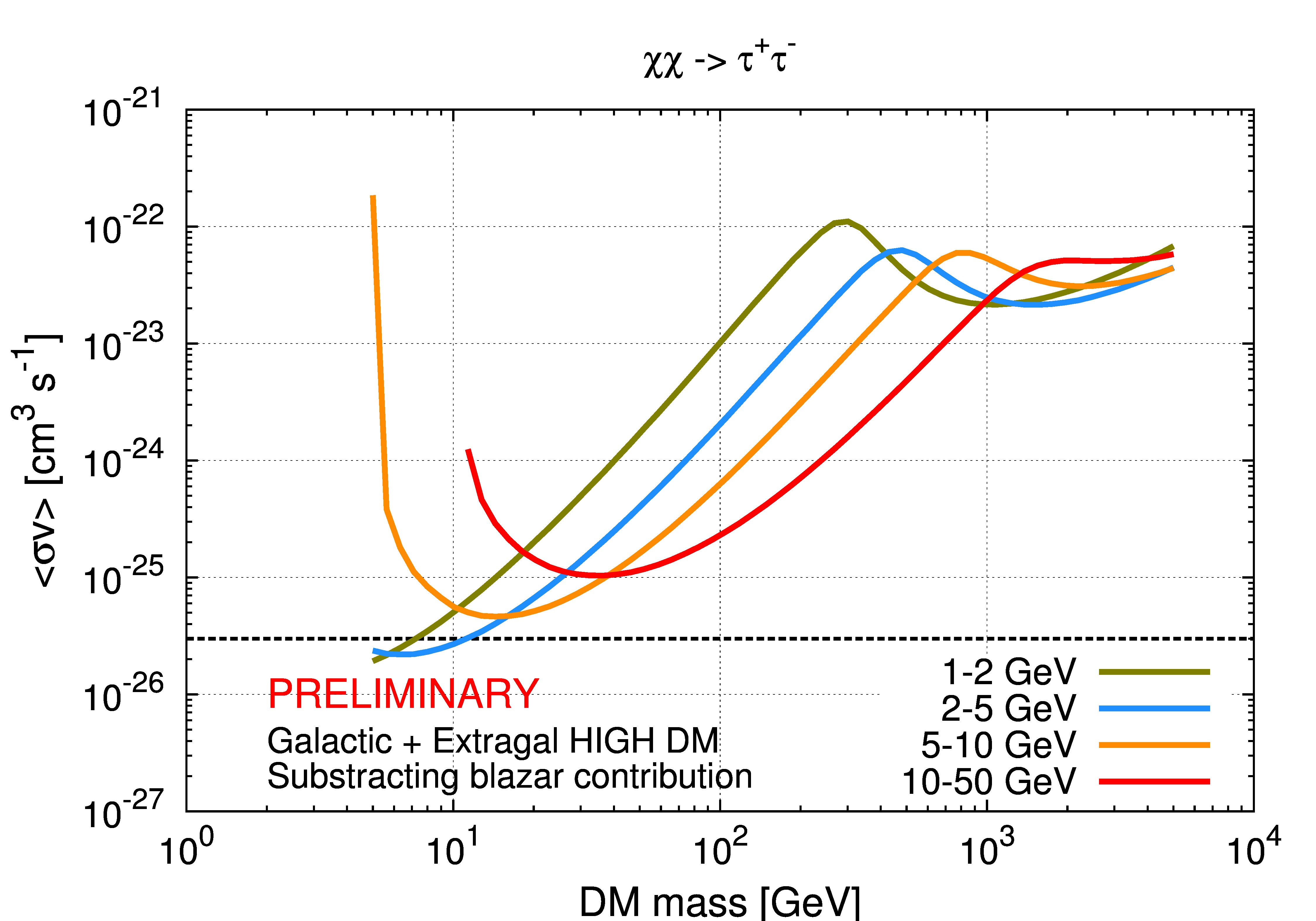}
\caption{$95\%$ CL limits on the annihilation cross section for the $\tau^+\tau^-$ channel.} \label{tau}
\end{figure*}

Figures \ref{bb} - \ref{tau} show the $95\%$ CL limits on the annihilation cross section for three different channels, $b\bar{b}$ quarks, $\mu^+\mu^-$, and $\tau^+\tau^-$ leptons, respectively.
The main uncertainty in the predictions obtained in~\citep{2013MNRAS.429.1529F} lies in the properties of  low-mass subhalos, below the mass resolution of the simulations. Different values of the "subhalo boost" strongly affect the prediction for the DM-induced gamma-ray intensity and its anisotropies. \citep{2013MNRAS.429.1529F} considered two benchmark scenarios for subhalos, assuming that the uncertainties can be modeled by changing the subhalo abundance: i) the LOW case, where halos are relatively poor in subhalos, according to the predictions of~\citep{5} and~\citep{6}; the constraints using this scenario are shown in the left-upper panel of figures \ref{bb} - \ref{tau}, and ii) the HIGH case, with large subhalo boosts, compatible with what was found by ~\citep{7,8}; the right-upper and lower panels in the figures \ref{bb} - \ref{tau} use this case. In the lower panels the predictions for the DM-induced APS (with a HIGH subhalo boost) are compared with the APS measured by Fermi-LAT once the contribution from blazars is subtracted. This represents the case where the most stringent constraints on $\left<\sigma v\right>$ are obtained.  We are currently updating the Fermi-LAT APS measurement using $\sim45$ months of data, we expect to improve the constraints presented in these proceedings after include the new measurement in the method shown in section~\ref{setting}.
\newpage

\newpage

\begin{acknowledgments}
This work was partially supported by the Spanish MINECO's Consolider-Ingenio 2010 
Programme under grant MultiDark CSD2009-00064. 

The work of GAGV was supported in part by MINECO under grants
FPA2009-08958, FPA2009-09017 and FPA2012-34694, 
and under the `Centro de Excelencia Severo Ochoa' Programme SEV-2012-0249, 
by the Comunidad de Madrid under grant HEPHACOS S2009/ESP-1473, 
and by the European Union under the Marie Curie-ITN program PITN-GA-2009-237920. 

GAGV thanks Caltech for hospitality during the completion of this work.

The work of TD was supported in part by the ANR project DMAstroLHC, ANR-12-BS05-0006-01.

JZ is supported by the University of Waterloo and the Perimeter Institute for
Theoretical Physics. Research at Perimeter Institute is supported by the
Government of Canada through Industry Canada and by the Province of Ontario
through the Ministry of Research \& Innovation. JZ acknowledges financial
support by a CITA National Fellowship.

The $Fermi$ LAT Collaboration acknowledges support from a number of agencies and institutes for both development and the operation of the LAT as well as scientific data analysis. These include NASA and DOE in the United States, CEA/Irfu and IN2P3/CNRS in France, ASI and INFN in Italy, MEXT, KEK, and JAXA in Japan, and the K.~A.~Wallenberg Foundation, the Swedish Research Council and the National Space Board in Sweden. Additional support from INAF in Italy and CNES in France for science analysis during the operations phase is also gratefully acknowledged.

\end{acknowledgments}
\newpage
\bigskip 

\end{document}